\documentclass[a4paper, 10pt,titlepage]{article}
\makeatletter                                               
\renewcommand{\@cite}[1]{#1}                                
\makeatother                                                
\newcommand{\onlinecite}[1]{\cite{#1}}                      
\newcommand{\supcite}[1]{{\tiny $^\textrm{\cite{#1}}$  }}

\newcommand{\rcite}[1]{Ref.~\onlinecite{#1}}               

\usepackage{epsfig}
\usepackage{psfrag}
\usepackage[latin1]{inputenc}

\newcommand{\fig}[1]{Fig. \ref{#1}}
\newcommand{\eq}[1]{Eq. (\ref{#1})} 
\newcommand{\df}[2]{\displaystyle\frac{#1}{#2}} 
\newcommand{\nop}{\hspace{-0.5mm}}

\begin{document} 
\twocolumn[\large {}]                                   
\setcounter{page}{1} 
\noindent
{\bf \Large Visualizing curved spacetime}\\[3mm] 
{\bf Rickard M. Jonsson}\\[2mm] 
{\it Department of Astronomy and Astrophysics,  
Physics and Engineering Physics, 
Chalmers University of Technology,
and G\"oteborg University,
412 96 Gothenburg, Sweden\\[2mm] 
}
E-mail: rico@fy.chalmers.se\\[2mm]
Submitted: 2003-04-23, Published: 2005-03-01\\
Journal Reference: Am. Journ. Phys. {\bf 73} 248 (2005)\\
\\
{\bf Abstract}. I present a way to {\it visualize} the concept of curved spacetime.
The result is a curved surface with local coordinate
systems (Minkowski Systems) living on it, giving the local directions of
space and time. 
Relative to these systems,
special relativity holds.
The method can be used to
visualize  gravitational time dilation, the horizon of
black holes, and cosmological models. 
The idea underlying the illustrations is first to 
specify a field of timelike
four-velocities $u^\mu$.
Then, at every point, one performs a coordinate
transformation to a local Minkowski system comoving with the given
four-velocity.  In the local system, the sign of the spatial part of the
metric is flipped to create a new metric of Euclidean signature. The new
positive definite metric, called the {\it absolute} metric, can be
covariantly related to the original Lorentzian metric. For the special
case of a 2-dimensional original metric, the absolute metric may be
embedded in 3-dimensional Euclidean space as a curved surface.

\section{Introduction} 
Einstein's theory of gravity is a geometrical theory and is well
suited to be explained by images. For instance the way a star affects
 the {\it space} around it can easily be
displayed by a curved surface.  The very heart of the theory, the {\it
curved spacetime}, is however fundamentally difficult to display using
curved surfaces. The reason is that the Lorentz signature gives us
negative squared distances, something that we never have on ordinary
curved surfaces.

However, we can illustrate much of the spacetime structure using flat
diagrams that include the lightcones. Famous examples of this are the
Kruskal and Penrose diagrams (see e.g. \rcite{dinverno}). Such pictures are valuable
tools for understanding black holes.

In this article I will describe a method that lets us visualize not just
the causal structure of spacetime, but also the scale (the proper
distances). It is my hope that these illustrations can be of help in
explaining the basic concepts of general relativity to a general
audience.

In exploring the possibilities of this method I use the language and
mathematics of general relativity. The level is that of teachers (or
skilled students) of general relativity. 

The resulting illustrations can however be used without any reference to mathematics
to explain concepts like 
gravitational time dilation, cosmological expansion, horizons
and so on. 

In Sec.~II, I give a brief introduction to the concept of curved
spacetime, using the method of this article, and consider a few examples of physical interest. 
This section presumes no knowledge of general relativity.
In Sec.~III, I explain the method underlying the illustrations thus far,
and also apply it to a black hole and a hollow star. 
In Secs.~IV-VIII, I present the general formalism,
demonstrate how to use it to produce embeddings and investigate the
geodesic properties of the formalism. 
These sections are of a more technical
character. In Secs.~IX-XIII, I apply the formalism
to various types of spacetimes. In Secs.~XIV-XVI, I relate this work to
other similar approaches, and comment on this
article. Secs.~XVII and XVIII include some pedagogical questions and
answers.

\section{Introduction to curved spacetime}
\begin{figure}[ht]
\begin{center} 
\epsfig{figure=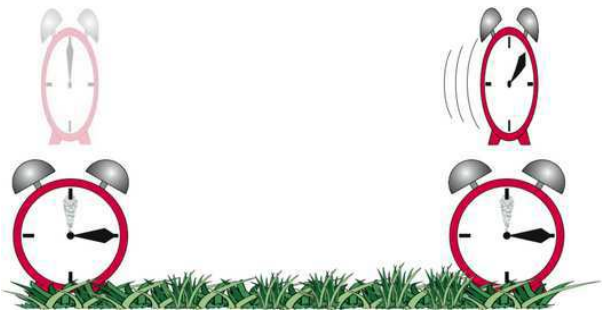,width=7.9cm}
\caption{A clock moving along a straight line. Relative to the clocks
at rest, the moving clock will tick more slowly.} 
\label{fig1} 
\end{center} 
\end{figure}

Consider a clock moving along a straight line. Special relativity
tells us that the clock 
will tick more slowly than the clocks at rest as illustrated in 
\fig{fig1}.

\begin{figure}[ht]
  \begin{center}
    \psfrag{dt}{$dt$}
   \psfrag{dx}{$dx$}
   \psfrag{t}{$t$}
   \psfrag{x}{$x$}
   \epsfig{figure=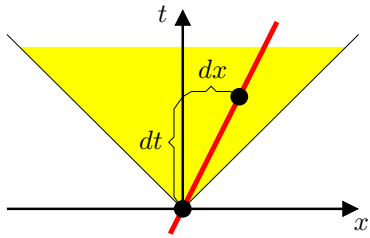,width=4.7cm}
    \caption{A spacetime diagram showing the worldline
    of the moving clock (the fat line). The two
    events we are considering are the black dots in the diagram. The
    shaded area is known as the lightcone.}
    \label{fig2}
  \end{center}
\end{figure}

Consider two events on the moving clock, separated by a time $dt$
and a distance $dx$, as seen relative to the system at rest. 
We can illustrate the two
events, and the motion of the clock in a {\it spacetime diagram}
as depicted in \fig{fig2}. Time is directed upwards in the diagram.
The motion of the clock corresponds to a {\it worldline} in
the diagram.

The proper time interval $d\tau$ is the time between the
two events according to the moving clock, which is given by\supcite{metric}
\begin{eqnarray}\label{hhh}
d\tau^2=dt^2 - \left( \frac{dx}{c} \right)^2.
\end{eqnarray}
Here $c$ is the velocity of light.
Note that in the limit as the speed of the clock 
approaches the speed of light we have $dx=c dt$, and thus from \eq{hhh} we have $d\tau=0$.
A clock moving almost at the speed of light will thus almost not tick at all
relative to the clocks at rest.

It is customary to choose the axes of the spacetime diagram in such a manner
that motion at the speed of light corresponds to a line that is
inclined at a $45^\circ$ angle relative to the axes of the
diagram. At every point in the diagram we can then draw a little
triangle, with a $90^\circ$ opening angle, known as a lightcone. The rightmost edge of the triangle corresponds
to a right-moving photon and the leftmost edge corresponds to a
left-moving photon. No material objects can travel faster than the
velocity of light, which means that the worldlines of objects must
always be directed within the local lightcone.

\subsection{Curved spacetime}
\begin{figure}[ht]
  \begin{center}
   \epsfig{figure=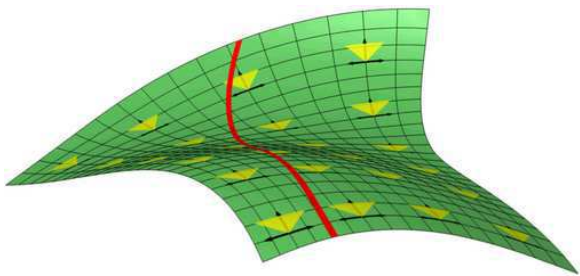,width=7.6cm}
    \caption{An illustration of curved spacetime using a curved surface
    with little Minkowski systems living on it. The curving line could
    be the worldline of a moving clock.}
	\label{fig3}
  \end{center}
\end{figure}

In general relativity we have a {\it curved} spacetime,
which we may illustrate by a curved surface with little
locally flat 
 coordinate systems, known as Minkowski
systems, living on it as illustrated in \fig{fig3}.

The little coordinate systems on the surface work precisely as the
spacetime diagram of \fig{fig2}. In particular the worldlines of
moving objects must always be directed within the local lightcone. 
To find out how much a
clock has ticked along it's winding worldline, we consider nearby
events along the worldline and sum up the $d\tau$'s we get using
\eq{hhh}, where $dt$ and $dx$ are the time and space separation
between the events as seen relative to the local Minkowski system.

\subsection{The spacetime of a line through a dense star}\label{hourglass}
\begin{figure}[ht]
\begin{center} 
\epsfig{figure=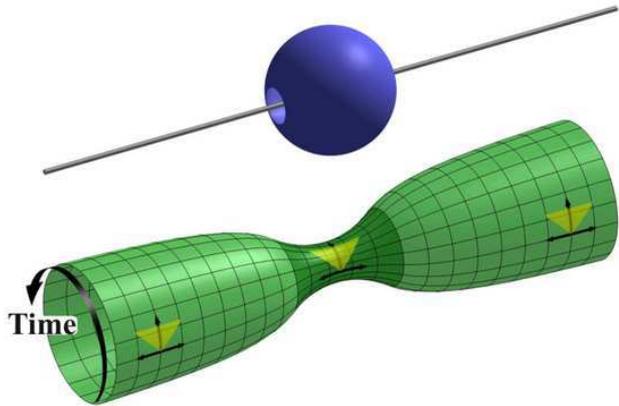,width=8.2cm}
\caption{A radial line through a very dense star, and an illustration
of the curved spacetime for that line.
Time is directed around the hourglass shaped surface.
Strictly speaking the surface should not close in on itself in the time
direction. Rather one should come to a new layer after one circumference
as on a paper roll.} \label{fig4} 
\end{center} 
\end{figure}

As a specific example let us consider the spacetime of a line through a
very dense star as depicted in \fig{fig4}.

The circles around the surface
correspond to fixed positions along the line through the star. The lines
directed along (as opposed to around) the surface correspond to fix
coordinate time (known for this particular case as Schwarzschild time). 

Consider now two observers, one at rest in the middle of the star
and the other at rest far to the left of the star. The worldlines of
these observers are circles around the middle and the left end of the
spacetime. 
Obviously the distance  measured around
the spacetime is shorter at the middle than at
the end. This means that the proper time (the experienced time) per
turn around the spacetime is shorter in the middle than at the end.
From this we may understand that time inside the
star runs slow relative to time outside the star. 

To be more specific consider the following scenario.
Let an observer far outside the star
send two photons, separated by a time corresponding to one lap, 
towards the center of the star.
The corresponding worldlines of the photons  will spiral around the
surface and arrive at the center of the star still separated by one lap.
The points where the photons arrive at the center will in this
illustration be the same, but they are
different points in spacetime because the surface is layered as in a
paper roll. 
Since the distance around the central part of the spacetime is smaller
than that towards the ends of the spacetime the observer in the center of the star will 
experience a shorter time  between the arrival of the two photons than 
the time between the emission of the two photons, as
experienced by the sender.
This effect is known as gravitational time dilation -- and is a consequence
of the shape of spacetime.

Alternatively we may note that the lines of constant coordinate time
are lying closer to each other in the middle of the spacetime surface
than at the ends. An observer inside the star will therefore observe that a local
 clock showing Schwarzschild coordinate time (synchronized with a
proper clock far outside the star) ticks much faster
than a clock measuring proper time within the star.
We may then understand that an observer inside the star will see the
Universe outside the star evolving at a faster rate than that experienced by an observer outside the star.

\subsection{Freely falling motion}
\begin{figure}[ht] \begin{center} 
\epsfig{figure=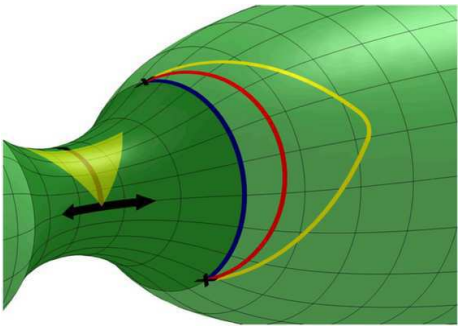,width=6.0cm} 
\caption{Three different worldlines connecting two fixed events. 
The middle worldline corresponds to the actual
motion of 
an object initially thrown radially
away from the star and then falling back towards the star. 
Of the three worldlines this has the largest integrated proper time.  }
\label{fig5} 
\end{center} 
\end{figure}

According to general relativity, an object thrown out radially from the
surface of the star, moving freely (so there is no air resistance for instance),
takes a path through spacetime such that the proper time
elapsed along the worldline of the object is {\it maximized}.
Consider then two events, at the surface of the star 
separated by some finite time only. It is easy to imagine
that a particle traveling between the two events will gain proper time
by moving out towards a larger embedding radius (where the
circumference is greater), before moving back to the second event. 
On the other hand it cannot move out too fast since
then it will move at a speed too close to the speed of light -- whereby
the internal clock hardly ticks at all, see \fig{fig5}. 

To predict the motion of an object that has been thrown out from the
star and returns to the same location after a specific amount of time, 
we can in
principle consider different pairs of events (as in \fig{fig5}),
find the worldline that maximizes the integrated proper time. This
trajectory corresponds to the motion that we are seeking.
Thus we can explain not only gravitational time dilation but also the motion
of thrown objects using images of the type shown in \fig{fig4} and \fig{fig5}.

\subsection{Cosmological models}\label{cosmosection} 
\begin{figure}[ht] \begin{center} 
\epsfig{figure=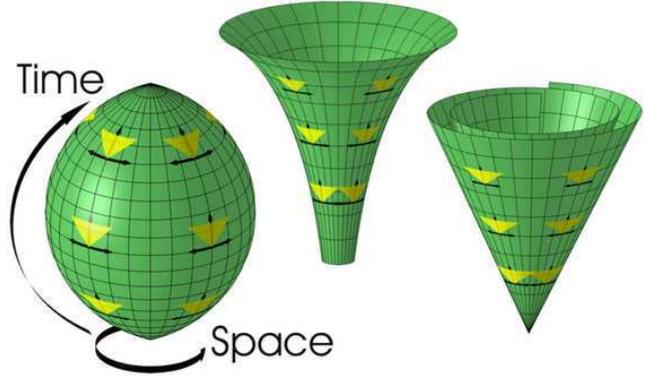,width=8.4cm} 
\caption{Schematic
spacetime cosmological models.} \label{fig6} \end{center} \end{figure}

We may use the the same technique that we employed in the previous
section to visualize the spacetimes
corresponding to various cosmological models (although we are
restricted to one spatial dimension). In \fig{fig6}, a few
images of such models are displayed.

\vspace{-0.3cm}

Notice that time is here directed along the surface and space is
directed around the surface. Just like before the local coordinate
systems, in which special relativity holds, gives the local
spatial and temporal distances.

The leftmost illustration corresponds to a Big Bang and Big Crunch
spacetime.  As we follow the spacetime upwards (i.e forward in time) the
circumference first increases and then shrinks. This means that space
itself expands and then contracts. The Big Bang is here just a point
on the spacetime -- where the spatial size of the universe was zero.
I will leave it to the reader to describe how space behaves in the two rightmost spacetimes.

Using Newtonian intuition one might think of 
the Big Bang as a giant fire cracker exploding at some
point in time. 
As the fire cracker explodes it sends out a cloud
of particles that expands at a great rate relative
to a {\it fixed} space. In Einstein's theory it is space itself that
expands due to the {\it shape} of spacetime. Also unlike in the fire
cracker view we cannot in general even
talk about a time before the Big Bang in Einstein's theory. 

As another application of the illustrations of \fig{fig6},  
consider a set of photon worldlines separated by some small spatial distance
shortly after the Big Bang in the leftmost Big Bang model.
The worldlines will spiral around the spacetime, 
always at $45^\circ$ to the local time axis. From this we may understand
that they will get further and further
separated as the circumference of the universe increases. Thinking of a photon as a set of
wave crests that are all moving at the speed of light, we then
understand that the wavelength of a photon will get longer and longer as the
universe grows larger. This effect is known as the cosmological red
shift. 
We can consider a similar scenario for the gravitational redshift by considering a photon moving along the  spacetime of the line through the dense star of \fig{fig4}.

\section{A simple method}\label{simple}
The idea allowing us to make a figure like \fig{fig4}, which is an
exact representation of the spacetime geometry, is simple.
Assume that we have a  two-dimensional, Lorentzian, time-independent and
diagonal metric: 
\begin{eqnarray} 
d\tau^2=g_{tt} dt^2 + g_{xx} dx^2.
\end{eqnarray} 
We then produce a new metric by taking the absolute
value of the original metric components:
\begin{eqnarray} 
d\bar{\tau}^2=|g_{tt}| dt^2 + |g_{xx}| dx^2.
\end{eqnarray} 
The new metric, called the absolute metric, is positive definite and
can be embedded in three dimensional Euclidean space as a surface of
revolution because $g_{tt}$ and $g_{xx}$ are independent of $t$.
For an observer with fixed
$x$, pure temporal and pure spatial distances will precisely correspond
to the absolute distances. There will thus be small Minkowski systems
living on the curved surface.    Analogous arguments hold if we have $x$
rather than
$t$-independence (as for the cosmological models).

\subsection{Black hole embedding} 
\begin{figure}[ht] \begin{center} 
\epsfig{figure=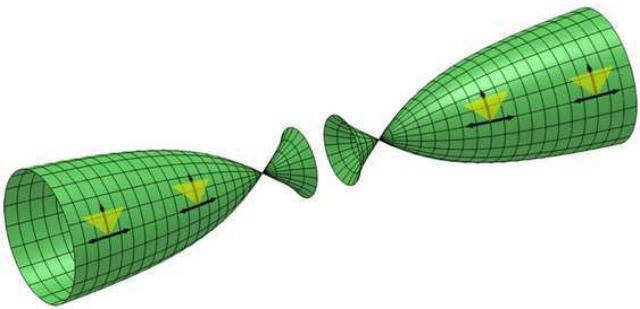,width=8.4cm} 
\caption{An embedding of the absolute spacetime of a central line
through a black hole.} \label{fig7} \end{center} \end{figure}

As another example of the visualization scheme outlined above, we consider the
line element of a radial line through a Schwarzschild black hole. An
embedding of the corresponding absolute metric is depicted in
\fig{fig7}.

\vspace{-0.3cm}

As before, the azimuthal angle corresponds to the Schwarzschild time.
The two points of zero embedding radius correspond to the horizon on
either side of the black hole. As we approach these points from the
outside, the time dilation becomes infinite. The trumpet-like regions
within these points lie within the horizon. Here moving {\it along} the
surface (as opposed to moving around the surface) corresponds to
timelike motion.

Photons, however, move at a $45^\circ$ angle relative to a purely azimuthal line,
both inside and outside of the horizon.  Studying a photon trajectory
coming from the outside and spiraling towards the point of zero
embedding radius, it is not hard to realize that it will take an
infinite number of laps (i.e. infinite Schwarzschild time) to reach that
point.

\begin{figure}[ht] \begin{center} 
\epsfig{figure=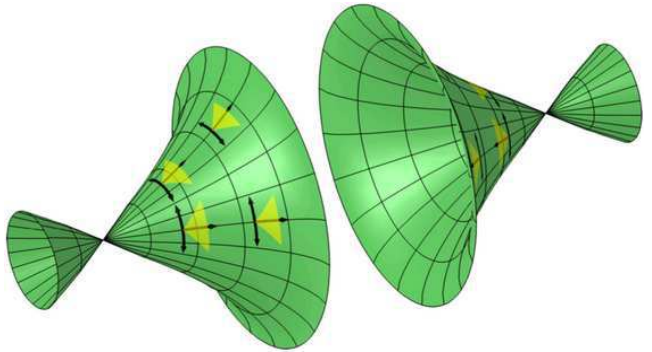,width=8.4cm} 
\caption{The absolute internal spacetime of a central line through a
black hole. Notice the direction of the lightcones. The singularity
lies in the (temporal) direction that the lightcones are opening up towards.
}
\label{fig8} \end{center}
\end{figure}

The singularity (where the spacetime curvature becomes infinite)
is not visible in the picture. While the distance as measured
along the internal trumpet from the horizon to the singularity is
finite (it has to be since we know that the proper time to reach the
singularity once inside the horizon is finite) the embedding radius
is infinite at the singularity. Thus, we cannot show the singularity
using this visualization. 
We can however come arbitrarily close by everywhere making the
embedding radius smaller. 
In \fig{fig8} we zoom in on the internal geometry.

\vspace{-0.3mm}
Note that the singularity is not a spatial point to which we may
walk. Once inside the horizon, the singularity lies in the {\it future} and it
is  impossible to avoid it -- just like it is impossible to avoid New Years
Eve.
 In this 1+1 dimensional scenario (inside the
horizon) the singularity is the time when space expands at an infinite rate.

Following a Schwarzschild time line (of fixed
azimuthal angle) inside the black hole corresponds to timelike
geodesic motion.
Imagine then two trajectories directed along two
such coordinate lines, starting close to the horizon and extending
towards the singularity. The corresponding two observers will be at
rest with respect to each other at the start (to zeroth order in the
initial separation between them). As they approach the singularity
they will however drift further and further apart in spacetime.
At the singularity, where the embedding radius is infinite, they will
be infinitely separated. We also know that the time it takes to reach
the singularity is finite. It is then easy to imagine that if we try
to keep the observers together, the force required will go to infinity
as we approach the singularity.  Hence, whatever we throw into a black
hole will be ripped apart as it approaches the singularity. Notice
however that there is no gravitational force in general
relativity. The shape of spacetime is in this case simply such that a
force is needed to keep things together, and in the end no force is
strong enough.

\subsection{Thin spherical crust}\label{crustsection} 
\begin{figure}[ht] \begin{center} 
\epsfig{figure=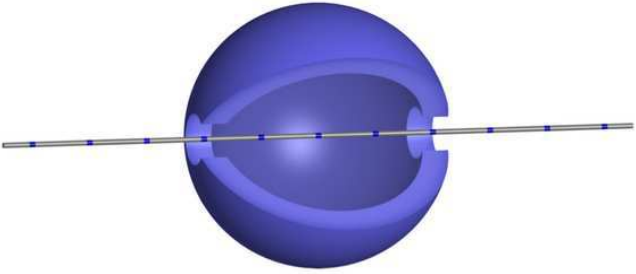,width=8.4cm} 
\caption{A line
through a thin crust of high mass. The wedge is cut out to obtain a
better view of the interior of the star.} \label{fig9} \end{center}
\end{figure}

As a pedagogical example, imagine a hollow massive star, with a
radial line through it, as illustrated in \fig{fig9}.

\begin{figure}[ht] \begin{center} 
\epsfig{figure=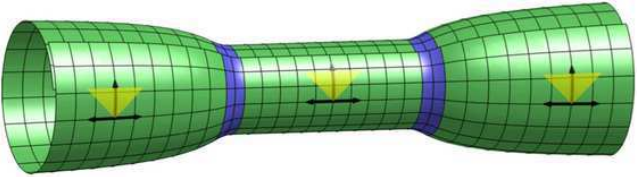,width=8.4cm} 
\caption{A schematic picture of the spacetime for a line through a
hollow star in the absolute scheme. Notice how, after one
circumference in time, we are really at a new layer.} \label{fig10}
\end{center} \end{figure}

We know from Birkhoff's theorem (see e.g. \rcite{weinberg}) 
that, assuming
spherical symmetry, the spacetime outside the crust will match the
external Schwarzschild solution. On the inside however, spacetime must
be Minkowski.\supcite{noblack} In \fig{fig10} the absolute
spacetime of the radial line is displayed.

If we were to cut out a square of the interior spacetime it would look
just like a corresponding square cut out at infinity. There is thus no
way that one, even by finite sized experiments (not just local
experiments) within the crust, can distinguish between being inside the
star or being at infinity. Even tidal effects are completely absent.

If we however were to open up a dialogue with someone on the outside, we
would find that the outside person would talk very fast, and in a high
pitched tone, whereas our speech would appear very slow and thick to the
outside person.

The  point that one  can illustrate  is that  we do  not have  to feel
gravity  for it  to  be there.  
Gravity  is not  about forces  pulling
things, it is about the fabric of space and time, and how
the different pieces of this fabric are  woven together.

\section{Generalization to arbitrary \\ spacetimes} 
The scheme outlined in the preceding section was specific for a particular
type of metric expressed in a particular type of coordinates. There is
however a way to generalize this scheme.

Given a Lorentzian spacetime of arbitrary dimensionality (although we
commonly will apply the scheme to two dimensions), the idea is first to
specify a field of timelike four-velocities denoted $u^\mu(x)$ (we will refer to spacetime velocities as four-velocities regardless of
dimensionality).
We then make a coordinate transformation to a local
Minkowski system comoving with the given four-velocity at every point.
In the local system we flip the sign of the spatial part of the metric
to create a new {\it absolute} metric of Euclidean signature. Notice
that the new metric will be highly dependent on our choice of
generating four-velocities. The absolute metric together with the
field of four-velocities contains all the information about the
original spacetime, and allows one to keep track of what is timelike
and what is not. We can always do the backwards
transformation and flip the local positive definite metric into a
Lorentzian (Minkowski) metric. 

Considering for example the black hole illustrations of the preceding
section, the {\it generators} (the worldlines tangent
to the field $u^\mu$) outside the horizon were simply those of the 
Schwarzschild observers at rest. Inside the
horizon the generators were the worldlines of observers for whom
$t=\textrm{const}$. Notice that the observers
located right outside the horizon have infinite
proper acceleration. It is then perhaps not surprising that the
resulting embedding is singular at the horizon. As we will see in
Sec.~\ref{freesec} we can better resolve the horizon by using the
worldlines of freely falling observers as generators.

\subsection{A covariant approach} \label{coap} 
We could carry out the scheme we just outlined
explicitly, doing coordinate transformations, flipping the sign of the
metric and transforming it back again. There is, however, a more
elegant method. We know that the absolute metric, from now on denoted
by $\bar{g}_{\mu\nu}$, is a tensor (as any metric),
and in a frame comoving with $u^\mu$ we have 
\begin{equation}
\bar{g}_{\mu\nu}\hspace{-0.1cm}=\hspace{-0.1cm} \left[ \hspace{-0.1cm}
\begin{array}{lllll} 1 \ \ 0  \ \   0 \ \   0                \\[-4mm] 0
\ \ 1  \ \   0 \ \   0                \\[-4mm] 0 \ \ 0  \ \   1 \ \   0 
              \\[-4mm] 0 \ \ 0  \ \   0 \ \   1 \end{array}
\hspace{-0.1cm} \right] \hspace{-0.1cm} = - \hspace{-0.1cm} \left[
\hspace{-0.1cm} \begin{array}{lllll} 1 \ \  0  \ \   0 \ \   0          
    \\[-4mm] 0 \   \textrm{-} 1  \ \  0 \ \  0               \\[-4mm] 0
\  \ 0  \   \textrm{-}1 \ \  0                \\[-4mm] 0 \  \ 0  \ \  0
\  \textrm{-}1 \end{array} \hspace{-0.1cm} \right] +2 \hspace{-0.1cm}
\left[ \hspace{-0.1cm} \begin{array}{lllll} 1 \ \ 0  \ \  0 \ \  0      
   \\[-4mm] 0 \ \ 0  \ \  0 \ \  0          \\[-4mm] 0 \ \ 0  \ \  0 \ \
 0          \\[-4mm] 0 \ \ 0  \ \  0 \ \  0 \end{array} \hspace{-0.1cm}
\right]. \end{equation} 
Adopting $(+,-,-,-)$ as the metrical signature
(as we will throughout the article), we realize that we must have:
\begin{eqnarray}\label{ppp}
\bar{g}_{\mu\nu}=-g_{\mu\nu} + 2  u_\mu u_\nu,   
\qquad \mbox{where}\qquad u_\mu=g_{\mu\nu}\frac{dx^\mu}{d\tau}. \label{hoppla}
\end{eqnarray} Notice that both sides of the equality are covariant
tensors that equal each other in one system, thus
the equality holds in every system.\supcite{ingemar}

For later convenience, we may also derive an expression for the
inverse absolute metric, defined by
$\bar{g}^{\mu\rho}\bar{g}_{\rho\nu}={\delta^\mu}_\nu$. 
By a contravariant argument, analogous to the covariant argument above, we
find that the inverse absolute metric is given by:
\begin{eqnarray}\label{inverse} \bar{g}^{\mu\nu}=-g^{\mu\nu} +
2  u^\mu u^\nu. 
\end{eqnarray} 
That this is indeed the inverse of the absolute metric can be verified
directly. It is a little surprising however that we get the inverse of
the new metric by raising the indices with the {\it original}
metric.\supcite{raise}

In the new metric, proper intervals will be completely
different from those in the original metric. Intervals as measured along a
generating congruence line will however be the same; these are
unaffected by the sign-flip. Using a bar to denote the four-velocity
relative to the absolute metric, it then follows that 
\begin{eqnarray}
u^\mu =\bar{u}^\mu, \qquad \qquad u_\mu =\bar{u}_\mu. 
\end{eqnarray} 
Using this in \eq{hoppla}, we immediately find
\begin{eqnarray}
g_{\mu\nu}=-\bar{g}_{\mu\nu} + 2  \bar{u}_\mu \bar{u}_\nu.
\end{eqnarray} 
Comparing with \eq{ppp}, we see that there is a perfect symmetry in going
from the original to the absolute metric, and vice versa.

\section{Freely falling observers as generators}\label{freesec} 
As a specific example of the absolute metric, we again consider the line
element of a radial line through a Schwarz\-schild black hole. We set
$c=G=1$, and introduce dimensionless coordinates, and proper intervals
\begin{eqnarray} 
x=\frac{r}{2M} \qquad t=\frac{t_{\textrm{\tiny
original}}}{2M} \qquad  \tau=\frac{\tau_{\textrm{\tiny original}}}{2M}.
\end{eqnarray} 
The line element then takes the form
\begin{eqnarray}
\label{sc} d\tau^2=\left(1-\frac{1}{x} \right) dt^2
-\left(1-\frac{1}{x} \right)^{-1} dx^2.
\end{eqnarray} 
As generators ($u^\mu$) we consider freely falling observers,
initially at rest at infinity. Using the squared Lagrangian formalism
(see e.g. \rcite{dinverno}) for the equations of motion, we readily find
the lowered four-velocity of the generating freefallers
\begin{eqnarray} 
u_\mu=\left(1,\frac{\sqrt{x}}{x-1}\right).
\end{eqnarray} 
The absolute metric is then according to \eq{hoppla}
\begin{equation} \bar{g}_{\mu\nu}= \left[ \begin{array}{ccccc}
1+\frac{1}{x} \ \ \ & \frac{2\sqrt{x}}{x-1} \\ \frac{2\sqrt{x}}{x-1} &
\frac{x(x+1)}{(x-1)^2} \end{array} \right] .
\end{equation} 
To make an embedding of this metric we are wise to first diagonalize
it by a coordinate transformation $t'=t+\phi(x)$. Letting
$\frac{d\phi}{dx}=\frac{\bar{g}_{tx}}{\bar{g}_{tt}}$ the line element
in the new coordinates becomes
\begin{eqnarray}\label{first}
d\bar{\tau}^2 = \left(1+\frac{1}{x}\right)  dt'^2 +
\left(1+\frac{1}{x}\right)^{-1}  dx^2. 
\end{eqnarray} 
This metric is easy to remember since by chance it is the
Schwarzschild metric with the minus signs replaced by plus signs
(except for the minus sign in the exponent).
Notice that nothing special happens with the metrical components at
the horizon ($x=1$). At the singularity ($x=0$) however, the absolute
metric is singular.

\begin{figure}[ht] \begin{center} 
\epsfig{figure=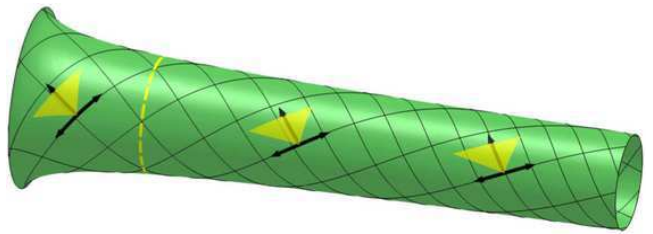,width=8.4cm} 
\caption{The absolute freefaller geometry. The dashed line is the
horizon. As before the singularity lies outside of the embedding
(towards the left).}
\label{fig11} \end{center} \end{figure}

To produce a meaningful picture of this geometry, we must include the
worldlines of the freely falling observers used to generate the
absolute geometry. Coordinate transforming of the trajectories to the new
coordinates $t',x$ 
can be done numerically. The result is depicted in \fig{fig11}.

Notice how the local Minkowski systems are twisted on the surface. The
horizon lies exactly where the generating worldlines are at a
$45^\circ$ angle to a purely azimuthal line.

Time dilation is now not solely determined by the local embedding
radius, but also by the gamma factor\supcite{gamma}
of the observer at rest relative to the
generating observer. For instance an observer at rest at the horizon
will be at a $45^\circ$ angle to the generating observer,
corresponding to an infinite gamma factor, and his clock will
therefore not tick at all during a Schwarzschild lap (one
circumference), thus being infinitely time-dilated.

Unlike the hour-glass type embeddings of Sec.~II~B, this
explanation of gravitational time dilation requires a basic knowledge
of special relativistic time dilation. 
However, unlike the illustration of \fig{fig7} where there is a cusp right at the horizon,
\fig{fig11} has the virtue of showing how passing the horizon is not at all dramatic
(locally). Spacetime is as smooth and continuous at the horizon as
everywhere else outside the singularity.

\section{Symmetry-preserving generators}\label{sympres} 
In this section we generalize
the scheme outlined in the preceding section to include arbitrary two-dimensional (2D)
metrics with a Killing symmetry,\supcite{killdef} for arbitrary
generators that preserve manifest Killing symmetry. In two dimensions,
the generating field $u^\mu$ can (since it is normalized) be specified
by a single parameter as a function of $x^\mu$. A parameter that 
is well suited to preserve the symmetries of the original metric is
the {\it Killing} velocity $v$. By this we mean the velocity that a
generator $u^\mu$ experiences for a Killing line (a worldline of
constant $x$). In other words it is the velocity of a point of
constant $x$ as seen by the generating observer. The absolute value of
this velocity will be smaller than one outside the horizon, and
greater than one inside the horizon.  Without loss of generality we
can assume that the original line element is of the form
Diag($g_{tt}(x),g_{xx}(x)$). The relation between $u^\mu$ and $v$ is
derived in Appendix \ref{umuv}. The result is
\begin{eqnarray}\label{umu} 
u^\mu=\pm \sqrt{\frac{g_{tt}}{1-v^2}} \left(
\frac{1}{g_{tt}},\frac{-v}{   \sqrt{-g_{xx}g_{tt}}  } \right).
\end{eqnarray} 
Using the lowered 
version of \eq{umu} in
\eq{hoppla} gives us the absolute metric as a function of the
parameter $v$. Making a coordinate transformation that diagonalizes
this metric, analogous to the diagonalization in the preceding
section,\supcite{diagonalize} yields after simplification
\begin{eqnarray}\label{gbardiag} \bar{g}'_{\mu\nu}= \left[
\begin{array}{cccc} g_{tt} \displaystyle \frac{1+v^2}{1-v^2} \qquad& 0
\\ 0 \qquad & \displaystyle -g_{xx} \frac{1-v^2}{1+v^2} \end{array}
\right]. 
\end{eqnarray}
Notice that if there is a
horizon present, where $g_{tt}=0$, we have also $(1-v^2)=0$. The
quotient of these two entities will remain finite and well defined,
given that $dv/dx\neq 0$ and $dg_{tt}/dx \neq 0$.

We see from \eq{gbardiag} that there is much freedom in choosing
$\bar{g}'_{tt}$. Since we can choose $v$ arbitrarily close to $1$,
both inside and outside of the horizon, we can everywhere make
$\bar{g}'_{tt}$ take an arbitrarily high value. Because the square root
of $\bar{g}'_{tt}$ is proportional to the embedding radius, there are
virtually no limits to what shape the curved surface 
can be given.
To interpret the embedded surface we need also the generating worldlines,
relative to the new (diagonalizing) coordinates. How these can be
found is derived in Appendix \ref{transvec}.

While the shape of the embedded surface depends strongly on the choice
of generators, the {\it area} is independent of this choice. 
This holds regardless of any assumed symmetries as is explained in
Appendix \ref{area}.

\section{Flat embeddings}\label{flatsec} 
\begin{figure}[ht] 
\begin{center} 
\psfrag{x=1}{$x=1$}
\psfrag{x=2}{$x=2$}
\epsfig{figure=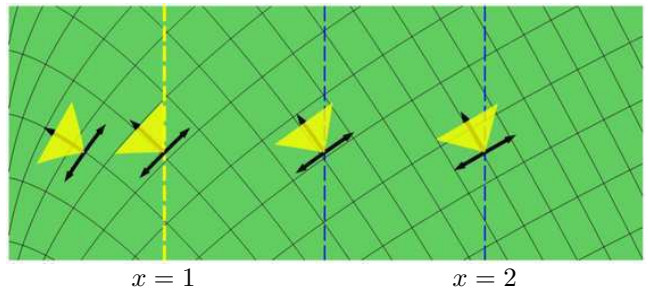,width=8.4cm} 
\caption{A flat
embedding of a Schwarzschild black hole. The radial parameter $x$ lies
in the interval [0.5, 2.5]. We could equivalently embed this geometry as
a cylinder. As we go further to the right (larger $x$), the lightcones
will approach pointing straight up.} \label{fig12} \end{center}
\end{figure}

Using \eq{gbardiag} and  assuming a time-symmetric and two-dimensional original
metric, we can produce an absolutely flat absolute metric. This we can
embed as a cylinder or a plane. We simply set $\bar{g}'_{tt}=C$, where
$C$ is some arbitrary positive constant. Solving for $v$ yields
\begin{eqnarray} v=\pm \sqrt{   \frac{C-g_{tt}}{C+g_{tt}}      }.
\end{eqnarray} 
As a specific example we consider a Schwarzschild original line
element. We choose  $v=0$ at infinity, corresponding to the
generating observers at infinity being at rest, which yields $C=1$. We
also choose the positive sign, corresponding to an in-falling observer
(on the outside) to find
\begin{eqnarray} 
v=\frac{1}{\sqrt{2x-1}}.
\end{eqnarray} 
This is a completely smooth function at the horizon. We see that it
remains real for $x\geq 1/2$.  For other choices of $C$ we can make
the inner boundary come arbitrarily close to the singularity. We
notice also that $\bar{g}'_{xx}=1/C$ and is thus also constant. This
means that the constant $x$-worldlines will be evenly spaced on the
flat surface. Also we may, from the expression for $v$, immediately
figure out how the local generator should be tilted relative to the
constant $x$-worldline on the flat surface. An embedding for this
particular case is displayed in \fig{fig12}.

Notice that for this visualization the curvature of spacetime is
manifested solely as a {\it twist} of the local Minkowski systems
relative to each other. As in the case of \fig{fig11}, the flat embedding
illustrates the smoothness of the spacetime around the horizon.

\section{Absolute geodesics} We know that the motion of particles in
free fall corresponds to trajectories that maximize the proper time.
Such trajectories can be found using the absolute scheme, as outlined
above. The fact that these trajectories are also {\it straight}, is
unfortunately a bit lost in this scheme. There are however ways to
manifestly retain at least parts of the original geodesic structure,
in the absolute metric. The net value of this discussion turns out to
be of more academic than pedagogical value. Below we therefore simply
summarize the results derived in the appendices.

\begin{itemize} \item Assuming an original 2D metric with a Killing
symmetry, we can demand that {\it some} given motion $x(t)$ should be
geodesic relative to the absolute metric. For example one can show that there exist 
generators such that outward-moving photons on a Schwarz\-schild
radial line follow absolute geodesics. For brevity the analysis of
this is omitted.

\item To investigate the general connection between the geodesic
structure of the original and the absolute metric, we derive a covariant
expression for the absolute four-acceleration in terms of Lorentzian
quantities. See Appendix \ref{covariant}.

\item Using the formalism of the preceding point we can show that if
the generators are geodesics with respect to the original metric 
they will also be with respect to the absolute metric, and vice versa. See
Appendix \ref{rjgeogen}. We also give an intuitive explanation for
this. 

\item In the preceding points we have seen how some parts of the
geodesic structure can be retained.  To completely retain the geodesic
structure, as is derived in Appendix \ref{equivalence}, we must have
\begin{eqnarray}\label{final} 
\bigtriangledown_\alpha u_\mu =0.
\end{eqnarray} 
At any {\it single} point in spacetime, this is easily achieved. We
just go to an originally freely falling system and in this system
choose $u^\mu=\frac{1}{\sqrt{g_{00}}} {\delta^\mu}_0$. Since in this
system the metric derivatives all vanish, so will the derivatives of
the generators.

For a normalized vector field $u^\mu$ to exist such that \eq{final}
holds everywhere, we must have a so called ultrastatic spacetime -- as
is derived in Appendix \ref{ultrastatic}. By ultrastatic we mean that
space may have some fixed shape, but there can be no time dilation.

\end{itemize}

\noindent We conclude that only to a limited extent can we, in the
absolute scheme, visualize Lorentz-geodesics as straight lines.
There are however other visualization methods that are better suited for this, 
as discussed in Sec.~\ref{other}.

\section{Charged black hole} 
\begin{figure}[ht] \begin{center}
\psfrag{x=1}{$x=1$}
\psfrag{x=2}{$x=2$}
\epsfig{figure=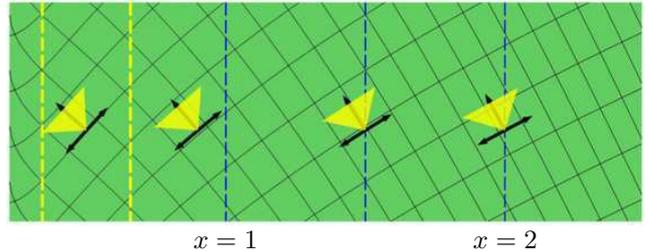,width=8.4cm} 
\caption{A flat
embedding of a Reissner-Nordstr\"om black hole. The dimensionless radial
coordinate $x$ lies in the range $[0.22, 2.5]$. The two internal
horizons are marked with the thicker dotted lines. The charge is chosen
so that $\beta=0.95$.} \label{fig13} \end{center} \end{figure}

All that we
have done so far for ordinary black holes, in the absolute scheme, can
also be done for charged black holes. The line element of a radial
line is then given by (see e.g. \rcite{gravitation}) 
\begin{eqnarray}\label{reis}
d\tau^2=\left(1-\frac{1}{x} + \frac{\beta^2}{4 x^2} \right) dt^2 -
\left(1-\frac{1}{x} + \frac{\beta^2}{4 x^2}\right)^{-1}
\hspace{-0.2cm}dx^2 \quad. 
\end{eqnarray} 
The dimensionless constant $\beta$ lies in the range $[0, 1]$ and
is proportional 
 to the charge of the black hole. Just as in
Sec.~\ref{flatsec}, we may find a flat absolute geometry for this line
element as depicted in \fig{fig13}.

\begin{figure}[ht] \begin{center} 
\epsfig{figure=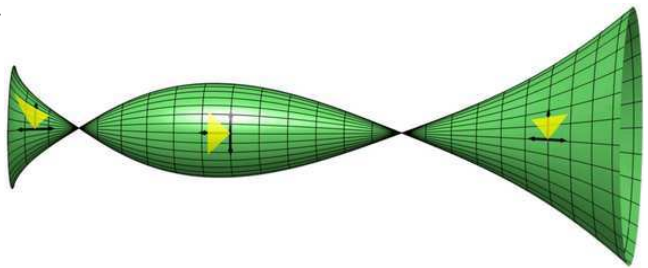,width=8.4cm} 
\caption{An
alternative representation of a Reissner Nordstr\"om black hole. Notice
the direction of the local Minkowski systems. Here $\beta=0.995$ and the
range is [0.425, 0.7].} \label{fig14} \end{center} \end{figure}

We see the classic three regions of the Reissner-Nordstr\"om solution.
Thinking of free particles taking a path that maximizes the proper
time we understand that a freely falling observer initially at rest in
the innermost region, will accelerate towards the inner
horizon. 
Actually this becomes clearer still if we form the absolute metric by simply taking the absolute
value of the original metrical components, as we did in
Sec.~\ref{simple}. This corresponds to having generators that are
orthogonal to the Killing field in the intermediate region, and
parallel to the Killing field outside this region. See \fig{fig14}.

We notice that the spacetime geometry of the region just inside the
inner horizon looks very much like the geometry just outside the outer
horizon. Knowing that it takes a finite proper time to reach the outer
horizon from the outside, we understand that it must take a finite
proper time (while infinite coordinate time) to reach the innermost
horizon from the inside. In the embedding there is however apparently
no region to which the trajectory may go after it has reached the
inner horizon. To resolve this puzzle, we must consider the extended
Reissner-Nordstr\"om solution. This is in principle straightforward, 
as will be briefly discussed at the end of
Sec.~\ref{kruskalsec}.

\section{Flat spacetime}\label{rindlersec} 
\begin{figure}[ht] \begin{center} 
\epsfig{figure=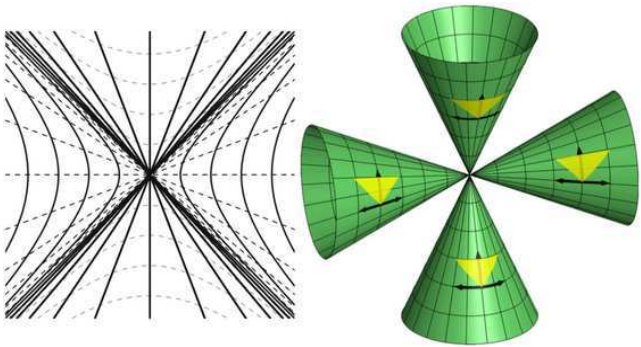,width=8.4cm} 
\caption{To the left: Minkowski spacetime with a certain set (as discussed in the main text)
of worldlines (the thick full drawn lines). To
the right: The corresponding absolute geometry embedding. Note that
the conical surfaces are not closed as one goes around in the space direction, but
rather they consist of very tightly rolled layers with no end.}
\label{fig15} \end{center} \end{figure}

The simplest
possible spacetime to which we may apply the absolute scheme, is flat
Minkowski in two dimensions. Choosing a field of generating
four-velocities that is constant, with respect to standard coordinates
($t,x$), the resulting absolute geometry is flat and can be embedded
as a plane. If we choose some more disordered field of four-velocities
we can however get an embedding with no apparent symmetries at
all. There is however another choice of generators that will produce a
regular surface,\supcite{seb} as is illustrated in \fig{fig15}.

In the right and left regions, of the Minkowski diagram, we have
chosen so called Rindler observers as generators.\supcite{rindler} In the
top and bottom regions we are using timelike
geodesics converging at the origin as generators. The universe as
perceived from this set of observers is known as
a Milne universe (in two dimensions).\supcite{rindler2}

It is obvious from the embedding that there is a (Lo\-rentzian) Killing
field directed around the conical surfaces. Imagining the corresponding
field in the diagram, we realize that it is in fact the Killing
field connected to continuous Lorentz transformations.

\section{Extended black hole}\label{kruskalsec} 
\begin{figure}[ht] \begin{center} 
\epsfig{figure=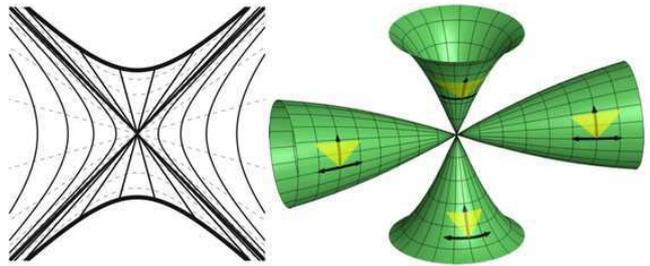,width=8.4cm} 
\caption{To the left: A Kruskal diagram of a maximally extended black
hole. To the right: an embedding of the absolute geometry with
generators at fixed radius in the exterior regions and at fixed
Schwarzschild time in the interior regions (the full drawn lines in
the diagram).} \label{fig16}
\end{center} \end{figure}

Notice the similarity between the Kruskal diagram of a maximally
extended Schwarzschild black hole (see e.g. \rcite{dinverno})
 and the Rindler diagram to the left in
\fig{fig15}. Having seen the interior and exterior regions of a (non--extended) black hole in the absolute
scheme (\fig{fig7}), we realize that we can also illustrate a maximally extended black hole
(\fig{fig16}).

While all symmetries are preserved in this picture, it is hard to see
how one can move between the different regions. Since the generators are
null at the horizons, making the absolute distance along these lines
zero, all the points along the null lines coming from the Kruskal origin
sit at the connecting point in the embedding. Thus a trajectory passing
one of the horizons in the diagram will pass through the connecting
point in the embedding. However, where it will end up is not evident from
the embedding alone. Through a more well behaved set of generators one
can remove this obscurity at the cost of loosing manifest symmetry, as
will be briefly discussed in Sec.~\ref{othersec}.

Having seen the absolute version of the extended
black hole, we can also figure out how the extended
Reissner-Nordstr\"om black hole must look. At all the cusps in the
embedding, four locally cone-like surfaces must meet. Otherwise, as is
apparent from Sec.~\ref{rindlersec}, the spacetime will not be complete.
I will leave to the readers imagination the specifics of
how to extend the Reissner-Nordstr\"om embedding depicted in \fig{fig14}.

\section{Other spacetimes}\label{othersec} 
So far in the embedding
examples, we have restricted ourselves to Lorentzian spacetimes with a
Killing symmetry and also to generators that manifestly preserve this
symmetry. The absolute scheme is however completely general. When
applying the scheme to the Kruskal black hole, we do not have to let
the generators be either parallel or orthogonal to the local Killing
field, as we did before. Instead we could for instance use geodesic
freefallers, originally at rest along a $t=0$ line in the standard
Kruskal coordinates. My best guess is that the corresponding embedding would
resemble a tortoise shell.

We can also consider spacetimes where there is no Killing symmetry. As
an example one could study a radial line through a collapsing thin
shell of matter. (Here there are local Killing fields but no global
Killing field.) As a first try, one might choose observers at fixed
$x$ as generators. Outside of the shell we would (via the Birkhoff
theorem) have a picture similar to \fig{fig7}. Inside the shell we
would have a flat (though it may be rolled up) surface, with straight
generating lines. Whether these two pieces can be put together in some
meaningful manner I have yet to discover. Maybe
one will find that another set of observers will be needed to join the
two spacetime regions together.

\section{Toy models} 
\begin{figure}[ht] \begin{center} 
\epsfig{figure=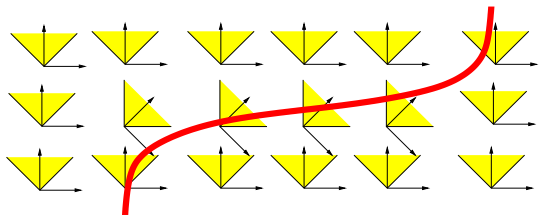,width=7cm}
\caption{A crude illustration of a toy model for warp drive}
\label{fig17} \end{center} \end{figure}

While we can use the absolute scheme to produce
pictures representing exact solutions to Einstein field equations, we
can also do the opposite. Suppose that we have a surface, say a plane,
and we specify an angle as a function of position on the
plane. Letting the angle correspond to the direction of the
generators, it is straightforward to find the
corresponding Lorentzian metric (just do the inverse transformation of
\eq{inverse}). We
may insert this metric into some program (say Mathematica) and let it
calculate the corresponding Einstein tensor and thus, through the
field equations, also the energy-momentum tensor.\supcite{energyconditions}
We may see the solution as a purely two-dimensional solution. 
Alternatively we may see it as a four-dimensional
solution assuming that we add two dimensions corresponding to
internally flat planes. In \fig{fig17} we see an example of such a toy
model spacetime.

From the look of the spacetime in \fig{fig17} we might call the propulsion mechanism
'twist drive' rather than warp drive. 
I will leave to the readers imagination to visualize a spacetime that
more deserves the name warp drive.

\section{Other methods} 
\begin{figure}[ht] \begin{center} 
\epsfig{figure=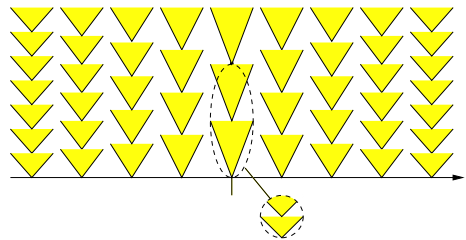,width=6cm}
\caption{A flat spacetime visualization of a radial line through a star.
The lightcones are everywhere, by definition, one proper time unit
high, and two proper length units wide. The dashed circle illustrates
that when we actually go to a specific region, the lightcones will appear
as they do at infinity.} \label{fig18} \end{center} \end{figure}

In this article we have seen how one may use
curved surfaces, with local Minkowski systems, to visualize for instance
gravitational time dilation. This can also be achieved using a flat
diagram, letting the space and time scales be encoded in the size (and
shape) of the local lightcones as depicted in \fig{fig18}.

The disadvantage with this technique is that it
is more abstract than the hourglass embedding
(\fig{fig4}). In the hourglass embedding there is a
shorter distance between two Schwarzschild time lines inside the star
than outside. From the flat lightcone model we must deduce this fact. Also one looses the visual
connection to the concept of curved spacetime.

The flat diagram technique however has the virtue of being extendable to 2+1
dimensions. The scheme outlined in this article can also be
used in 2+1 dimensions, but to produce a faithful image
we would need a flat {\it absolute} spacetime. Then we could embed
little lightcones of constant opening angle and size. To demand a
Euclidean absolute spacetime is however quite restrictive, and it
seems better to allow the lightcones to vary in
apparent width and height. 
I will leave to the reader's imagination
how this technique could be applied to visualize warp drive, rotating
black holes, the Big Bang and so on.

\section{Comparison to other work}\label{other}
There are, to the author's
knowledge, three other distinctly different techniques of visualizing
 curved spacetime using embedded surfaces.

Marolf\supcite{marolf}  presents a way of embedding a
two-dimensional Lorentzian metric in a 2+1 dimensional Minkowski spacetime
(visualized as a Euclidean 3-space).

L. C. Epstein\supcite{epstein}  presents a popular scientific
visualization of general relativity. The underlying theory rests on the
assumption of an original time independent, diagonal, Lorentzian line
element.  Rearranging terms in this line element one can get something
that looks like a new line element, but where the proper time is now a
coordinate. The 'space-propertime' can be embedded as a curved
surface, from which many spacetime properties can be deduced.

\begin{figure}[ht] \begin{center}
\psfrag{T}{Time}
\psfrag{ime}{}
\epsfig{figure=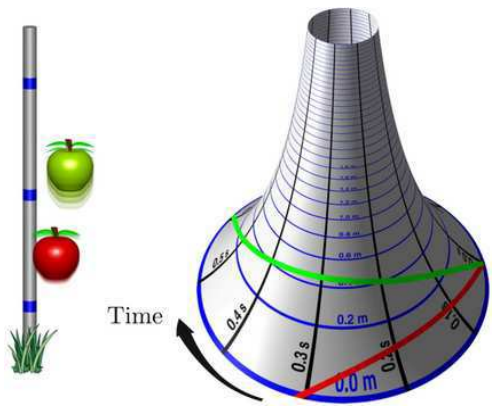,width=6.4cm} 
\caption{Illustration of  how straight lines in a curved spacetime can
explain the motion of upwards-thrown apples. The lines can
be found using a little toy car that is rolled straight ahead on the
surface.}
\label{fig19} \end{center}
\end{figure}

In a previous article\supcite{rickard}, I assumed a time independent
Lorentzian line element. I then found another line element, also time
symmetric, that is positive definite and geodesically equivalent to the
original line element. The resulting geometry can be embedded as a
curved surface as in \fig{fig19}.
The method can be used to explain straight lines in a curved
spacetime, the meaning of forces as something that bends spacetime 
trajectories, etc.

Each of the three techniques outlined here together with the absolute scheme
of this article has  different virtues (and drawbacks). 
Depending on the audience they can all be used to
explain aspects of the theory of general relativity.

\section{Comments and conclusions} The absolute
scheme as presented in this article, can be applied to {\it any}
spacetime, giving a global positive definite metric. Together with the
generating field of four-velocities, it carries complete information
about the original Lorentzian spacetime.

While there are mathematical applications of this scheme (see
\rcite{hawking}), we have here focused on its pedagogical virtues. At
the level of students of relativity, a study of
the mathematical structure itself may have some pedagogical virtue. In
particular it is instructive to see an alternative representation of
the shape of spacetime.

The main pedagogical virtue is however that, applied to two-dimensional
spacetimes, the absolute scheme enables us to make {\it embeddings}
that illustrate the meaning of a curved spacetime.

While such an embedding is not unique, due to the freedom in choosing
generators as well as the freedom of the embedding, it gives a
completely faithful image of the true spacetime geometry. With only a
basic understanding of Minkowski systems, a complete knowledge of 
the embedded parts of the Lorentzian geometry can
be deduced from the surface.  For instance we can figure out how much
a clock has ticked along a certain path, or what path a thrown
apple will take.

Also, many applications require no knowledge at all of special
relativity. 
We do not need to mention
Minkowski systems or proper times 
to give a feeling for how
geometry can explain how time can run at different rates at different
places, or how space itself can expand.
Most importantly we emphasize 
the point of view that gravity, according to general relativity, is about
{\it shapes} -- not forces and fields.

\section{Questions for students of \\ general
relativity} Here are a handful of questions regarding applications of
the absolute scheme. The answers are given in the next section.
\begin{enumerate} \item The hour-glass shaped embedding of \fig{fig4}
illustrates a spacetime where time ``runs slower'' in a local region.
How would a corresponding  embedding look that illustrates how time can
run faster in a local region?

\item Can you, using the technique of this article, illustrate a
two-dimensional spacetime that is closed in space and time and with no
vertices (by a vertex we mean a point from which the Minkowski systems
point either outward or inward)?

\item Can a spacetime of the type specified in the preceding question
be flat?

\item In exam periods students often need more time to study. Consider
as a primary spacetime a flat plane with uniformly directed Minkowski
systems. How would you alter this spacetime to ensure that there is
sufficient time to study? Include the worldline of the student in need,
as well as the worldline of the teacher bringing the exam.

\item Imagine an upright-standing cylindrical surface, with
upward-directed Minkowski systems. The all-famous experiment where one
twin goes on a trip and later returns to his brother, can be illustrated
by two worldlines on the cylinder, one going straight up and the second
going in a spiral around the cylinder (intersecting the first one
twice). This scenario differs from the standard one in that no
acceleration was needed by either twin for them to still reunite. The
same question applies however. Will the twins have aged differently?

\end{enumerate} 

\section{Answers to students questions} Here are
(the) answers to the questions of the preceding section.
\begin{enumerate}

\item Instead of a dip in the hour-glass shaped embedding (decrease of
the  radius towards the middle), we have a bulge (increase of
 radius towards the middle).

\item Yes. For instance a torus, with the Minkowski systems directed
along the smaller toroidal circumference.

\item Yes. Make a tube out of a paper by taping two opposite ends
together. Flatten the tube, preferably so that the tape is a bit away
from the two folds that emerges. Roll the flattened tube, so that the
tape describes a complete circle and connect the meeting paper ends by
some more tape. If the local Minkowski systems on this shape are given a
uniform direction, the corresponding Lorentz-geometry will be flat.

\item Make a sufficiently high and steep bump in the plane, while
keeping the direction of the Minkowski systems (as seen from above the
former plane). Make sure that the student's trajectory passes straight
over the peak, while the teacher's trajectory misses it.

\item Oh yes. What time the twins experience is determined by their
respective spacetime trajectories. If the trajectory of the traveling
twin is tilted almost as much as a photon trajectory, he will have aged
very little compared to his brother. There is also an article\supcite{twin} 
that deals with this thought-experiment.

\end{enumerate}

\appendix

\section{Finding $u^\mu$ as a function of $v$}\label{umuv} 
\begin{figure}[ht] \begin{center}
\psfrag{u}{$u^\mu$}
\psfrag{v}{$v^\mu$}
\psfrag{x}{$\xi^\mu$}
\epsfig{figure=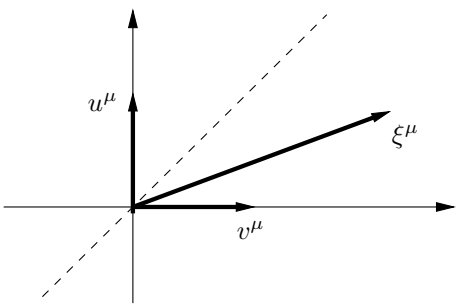,width=6cm}
\caption{The Killing field in local coordinates comoving with $u^\mu$.}
\label{fig20} \end{center} \end{figure}

Here is a
derivation of the general expression for $u^\mu$ as a function of the
Killing velocity $v$. Let us define $v^\mu$ as a vector perpendicular to
$u^\mu$, normalized to $-1$. Also we denote the Killing field by $\xi
^\mu$. The vectors as seen relative to a system comoving with $u^\mu$ are
displayed in \fig{fig20}.

We have then 
\begin{eqnarray}\label{ettan} \xi^\mu=(u^\mu + v v^\mu) K.
\end{eqnarray} 
Here the variable $K$ may take positive or a negative values.
Contracting both sides with themselves, we solve for $K$ to find
\begin{eqnarray}\label{tvaan} K=\pm \sqrt{   \frac{\xi^\alpha
\xi_\alpha}{1-v^2}}.
\end{eqnarray} 
The orthogonal vector $v^\mu$ can be
expressed via
\begin{eqnarray}\label{trean}
v^\mu=\frac{1}{\sqrt{g}}\epsilon ^{\mu\rho} g_{\rho \alpha} u^\alpha
\qquad \textrm{where} \qquad \epsilon^{\mu\rho}= \left( \nop \nop \nop 
\begin{array}{cccc} 0 \quad& \textrm{-}1\\[-2mm] 1  \quad& 0\\
\end{array} \nop \nop \nop \right). 
\end{eqnarray} 
Here $g=-\textrm{Det}(g_{\mu\nu})$. Through this definition $v^\mu$ is within
$180^\circ$ clockwise from $u^\mu$ (looking at the coordinate plane from
above, assuming $t$ up and $x$ to the right). Inserting \eq{trean} into
\eq{ettan}, using \eq{tvaan}, we readily find 
\begin{eqnarray} \left(
{\delta^\mu}_\nu + \frac{v}{\sqrt{g}} \epsilon ^{\mu\rho} g_{\rho \nu} 
\right) u^\nu= \pm \sqrt{\frac{1-v^2}{\xi^\alpha \xi_\alpha}}.
\end{eqnarray} 
This is a linear equation for $u^\mu$ that can easily be
solved. For the particular case of $g_{\mu\nu}=
\textrm{Diag}(g_{tt},g_{xx})$ and $\xi^\mu=(1,0)$ we find
\begin{eqnarray}\label{umuapp} u^\mu=\pm \sqrt{\frac{g_{tt}}{1-v^2}}
\left( \frac{1}{g_{tt}},\frac{-v}{   \sqrt{-g_{xx}g_{tt}}  } \right).
\end{eqnarray} 
So now we have a general expression for the generating four-velocity,
expressed in terms of the Killing velocity $v$. The $\pm$ originates
from the $\pm$ in the previous expression for $K$.

\section{Vector transformation by diagonalization}\label{transvec} 
Under the diagonalization of the
absolute metric (as performed in Sec.~\ref{sympres}) a general vector transforms according to
\begin{eqnarray}\label{transform}
q'^\mu=\left(q^t+\frac{\bar{g}_{tx}}{\bar{g}_{tt}} q^x, {}{}{}
q^x\right). 
\end{eqnarray} 
Using the lowered 
version of \eq{umu} and the
definition of the absolute metric, \eq{hoppla}, we find after
simplification 
\begin{eqnarray}\label{factor}
\frac{\bar{g}_{tx}}{\bar{g}_{tt}}=\frac{2v}{1+v^2}
\frac{-g_{xx}}{\sqrt{-g_{tt} g_{xx}}}. 
\end{eqnarray} 
For the particular case of $q^\mu=u^\mu$, using \eq{umu}, we find
after simplification
\begin{eqnarray}\label{umustar} 
u'^{\mu}=\pm
\sqrt{\frac{g_{tt}}{1-v^2}}\left(\frac{1}{g_{tt}}\frac{1-v^2}{1+v^2},
\frac{-v}{\sqrt{-g_{xx}{g_{tt}}}}\right). 
\end{eqnarray} 
This expression can be used to find the generating lines in the new
coordinates. We simply integrate $u'^t$ and $u'^x$ numerically, with
respect to the parameter $\tau$, to find $t'(\tau)$ and
$x(\tau)$. These lines can then easily be mapped to an embedding of
the absolute geometry.

\section{Regarding the absolute area}\label{area} 
Consider a small square in the coordinates we are
using. Then consider two different choices of generators, ${u_1}^\mu$
and ${u_2}^\mu$. Assume also that the coordinate square is small
enough that the generating fields (as well as the metric) can be
considered constant within the surface. We denote the absolute area of the
coordinate square for the two representations by $dA_1$ and
$dA_2$. Knowing that the absolute area is independent of the choice of
coordinates we may evaluate each area in the local Minkowski systems
${x_1}^\mu$ and ${x_2}^\mu$, comoving with the corresponding
generator. In these systems the original square will be deformed, but
the absolute area will exactly equal the coordinate area. Since the
two Minkowski systems are related via the Lorentz transformation,
which preserves coordinate areas, we know that the coordinate areas
are equal and thus also $dA_1=dA_2$. Because the argument applies to
arbitrary pairs of generators, it follows that the absolute area is
independent of the choice of generators.

In $N$ dimensions we can, by a completely analogous argument, show
that the $N$-volume of the absolute metric is independent of the
choice of generators.

\section{Covariant relation for the absolute
four-acceleration}\label{covariant} 
First we derive an expression for
a general absolute four-velocity $\bar{q}^\mu$, in terms of Lorentzian
quantities \setlength\arraycolsep{1pt}
\begin{eqnarray}\label{fourvelocity} 
\bar{q}^\mu&=&\df{dx^\mu}{d\bar{\tau}}=\df{dx^\mu}{d\tau} \df{d\tau}{d\bar{\tau}} \nonumber\\[2mm]
&=&q^\mu  \sqrt{ \df{g_{\mu\nu} dx^\mu dx^\nu}{-g_{\mu\nu} dx^\mu
dx^\nu +2 u_\mu u_\nu dx^\mu dx^\nu}} \nonumber\\[3mm] 
&=&q^\mu  \sqrt{\df{1}{-1+2 u_\mu u_\nu \df{dx^\mu}{d\tau} \df{dx^\nu}{d\tau}}  } \nonumber \\ 
&=&q^\mu   \df{1}{\sqrt{2 (u_\mu q^\mu)^2-1}} {\ }. 
\end{eqnarray}
Notice that choosing $q^\mu=u^\mu$ yields $\bar{u}^\mu=u^\mu$ as we
realized before.

To covariantly relate the absolute four-acceleration to the Lorentzian
quantities we need an expression for the absolute affine connection, i.e.
the affine connection for the absolute metric 
\begin{eqnarray}
\bar{\Gamma}^\mu_{\alpha\beta}=\frac{1}{2} \bar{g}^{\mu\rho}
\left(\partial_\alpha \bar{g}_{\rho\beta} + \partial_\beta
\bar{g}_{\rho\alpha} - \partial_\rho \bar{g}_{\alpha\beta} \right).
\end{eqnarray} 
Using the corresponding definition of the original affine
connection, together with the expressions for the absolute metric and
its inverse given by Eqs.~(\ref{hoppla}) and (\ref{inverse}), we can write this as
\setlength\arraycolsep{1pt} 
\begin{eqnarray}\label{affine}
\bar{\Gamma}^\mu_{\alpha\beta}&=&
\Gamma ^\mu_{\alpha\beta} - 2 u^\mu u^\rho  
\Gamma_{\rho\alpha\beta} +\left(\textrm{-}g^{\mu\rho} + 2 u^\mu
u^\rho\right)   \nonumber \\ 
&& \quad \times \left(\partial_\alpha
(u_\rho u_\beta) + \partial_{\beta} (u_\rho u_\alpha) - \partial_{\rho}
(u_\alpha u_\beta) \right) \label{gamma}. 
\end{eqnarray}
Now we evaluate $\frac{\bar{D}\bar{q}^\mu}{\bar{D}\bar{\tau}}$ in an
{\it originally} freely falling system where the original affine
connection vanishes. Setting all unbarred derivatives to their
covariant analog and using the definition of covariant
derivatives, Eqs.~(\ref{fourvelocity}) and (\ref{affine}) we obtain
\setlength\arraycolsep{1pt}
\begin{eqnarray}\label{jiha}
\frac{\bar{D}\bar{q}^\mu}{\bar{D}\bar{\tau}}=&&\frac{1}{\sqrt{2 (u_\mu
q^\mu)^2-1}} \frac{D}{D\tau} \left(\frac{q^\mu}{\sqrt{2 (u_\mu
q^\mu)^2-1}}\right)   \nonumber\\ 
&&+2 \frac{q^\alpha q^\beta}{2 (u_\mu q^\mu)^2-1} 
\left( -g^{\mu\rho}+2 u^\mu u^\rho  \right)   \nonumber \\ 
&&\times (  u_\beta \bigtriangledown_\alpha u_\rho + u_\rho \bigtriangledown_\alpha
u_\beta - u_\beta \bigtriangledown_\rho u_\alpha    ).
\end{eqnarray} 
Here we have a manifestly covariant relation. We can
expand this expression, and simplify it somewhat using
\begin{eqnarray}\label{norm} u^\mu u_\mu=1 \qquad \quad u^\mu
\bigtriangledown_\alpha u_\mu=0 \qquad \quad k\equiv u^\mu q_\mu.
\end{eqnarray} 
The first two relations follow directly from the normalization of
$u^\mu$, and the latter is a definition introduced for
compactness. The resulting expansion is given by
\setlength\arraycolsep{1pt} \begin{eqnarray}\label{jiha2}
\frac{\bar{D}\bar{q}^\mu}{\bar{D}\bar{\tau}}=&&\frac{2}{2 k^2-1}  \nonumber \\ 
\Big[&& \frac{1}{2} \frac{D q^\mu}{D\tau} - \frac{k
q^\mu}{2k^2-1} (q^\rho q^\sigma \bigtriangledown_\sigma u_\rho +
u_\rho \frac{D}{D\tau} q^\rho) \nonumber 
\\&& - k q^\alpha \bigtriangledown_\alpha u^\mu + u^\mu q^\beta q^\alpha
\bigtriangledown_\alpha u_\beta  \nonumber
\\&& + kq^\alpha \bigtriangledown^\mu
u_\alpha - 2k u^\mu q^\alpha u^\rho \bigtriangledown_\rho u_\alpha \Big]. 
\end{eqnarray}

\section{Geodesic generators}\label{rjgeogen} Consider a trajectory that
everywhere is tangent to the generating field so that $q^\mu=u^\mu$. Also,
assume that the generators are geodesics $\frac{D u^\mu}{D\tau}=0$, or
equivalently $u^\rho\bigtriangledown _\rho u^\mu=0$. Using the
normalization relation $u^\mu u_\mu=1$, from which it follows that $u^\mu
\bigtriangledown_\alpha u_\mu=0$, we immediately see that \eq{jiha}
reduces to 
\setlength\arraycolsep{1pt} 
\begin{eqnarray}\label{necsuf}
\frac{\bar{D}\bar{u}^\mu}{\bar{D}\bar{\tau}}=0. 
\end{eqnarray} 
Thus if the original generators are geodesic then they are geodesic also
relative to the absolute metric. Through the perfect symmetry in
transforming from the absolute to the Lorentzian metric and back, we
have derived implicitly that if the absolute generators are geodesics
they will also be geodesics in Lorentzian spacetime. We conclude that
{\it if and only if} the original generators are geodesic then they will
be geodesic in the absolute spacetime.

To get an intuitive feeling for the
result we just derived consider a straight  generating line (in the
absolute sense) on an embedded surface. Any
small deviation from this line will introduce negative contributions to
the proper time. A rigorous argument is that an infinitesimal variation
of a trajectory (with fixed endpoints), around a straight generating line,
will to first order in the variation parameter not affect the {\it
absolute} length of the trajectory. Also we know that the Lorentzian
distance along a trajectory is shorter than or equal to the absolute
distance (the equality holds if and only if we follow a generator).
Hence we cannot gain proper time to first order in the variational
parameter as we vary the trajectory. Thus the Lorentz proper time is
maximized by the original trajectory.

On the other hand, if the absolute generating line is curving relative
to the surface, it seems plausible that we could gain proper time by
taking a path on the {\it outside} of the curving generating line. Thus
a non-geodesic absolute generator would imply a non-geodesic Lorentzian
generator. The result that the generators are absolute geodesics if and
only if they are Lorentzian geodesics, is thus intuitively
understandable.

\section{Deriving necessary and suffic\-ient conditions for geodesic
\\ equivalence}\label{equivalence} To investigate if it is possible to
completely retain the original geodesic structure in the absolute
metric, we set $\frac{\bar{D}\bar{q}^\mu}{\bar{D}\bar{\tau}}=0$ and
$\frac{D q^\mu}{D\tau}=0$ in \eq{jiha2}. The resulting equation is given
by 
\begin{eqnarray}\label{jiha3} 0=&&- \frac{k q^\mu}{2k^2-1} q^\rho
q^\sigma \bigtriangledown_\sigma u_\rho - k q^\alpha
\bigtriangledown_\alpha u^\mu + u^\mu q^\beta q^\alpha
\bigtriangledown_\alpha u_\beta \nonumber \\&& + kq^\alpha
\bigtriangledown^\mu u_\alpha - 2k u^\mu q^\alpha u^\rho
\bigtriangledown_\rho u_\alpha. 
\end{eqnarray} 
If this equation is to hold for {\it all} directions, $q^\alpha$, it
must hold for the particular case $q^\alpha=u^\alpha$. Inserting this
and using the normalization of $u^\mu$, only the second term survives
\begin{eqnarray} u^\alpha \bigtriangledown_\alpha u^\mu =0.
\end{eqnarray} 
Thus it is {\it necessary} to have geodesic generators to get all the
geodesics 'right'. Assuming the generators to be geodesic -- the last
term in \eq{jiha3} dies. Multiplying the remaining four terms by
$q_\mu$, we are after simplification left with another necessary
constraint
\begin{eqnarray} \left(\frac{-k}{2k^2-1}+k\right) q^\alpha
q^\mu \bigtriangledown_\alpha u_\mu=0. 
\end{eqnarray} 
The expression within the parenthesis is zero if and only if $k=\pm
1$. Assuming a future-like convention on both $u^\mu$ and $q^\mu$ we
cannot have a negative $k$, and $k=1$ corresponds uniquely to
$u^\mu=q^\mu$, a direction that we already considered. Thus,
the expression outside the parenthesis must vanish. For this to hold for {\it all}
directions $q^\alpha$, we must have 
\begin{eqnarray}
\bigtriangledown_\alpha u_\mu=-\bigtriangledown_\mu u_\alpha.
\end{eqnarray} 
Using this necessary antisymmetry in \eq{jiha3} we are left with
\begin{eqnarray}\label{temp} 
0=&&- k q^\alpha \bigtriangledown_\alpha
u^\mu + kq^\alpha \bigtriangledown^\mu u_\alpha. 
\end{eqnarray} 
Lowering this with $g_{\mu\nu}$ and using the necessary antisymmetry
again, we obtain
\begin{eqnarray} q^\alpha \bigtriangledown_\alpha u_\nu
=0.
\end{eqnarray} 
For this in turn to hold for {\it all} $q^\alpha$ it is
necessary to have $\bigtriangledown_\alpha u_\mu =0$. This also
immediately satisfies the above necessary constraints on antisymmetry
and generator geodesics. That it is also {\it sufficient} for geodesic
equivalence follows directly from \eq{jiha3}. Thus the absolute metric
will be geodesically equivalent to the original one, if and only if
\begin{eqnarray}\label{finalappendix} 
\bigtriangledown_\alpha u_\mu =0.
\end{eqnarray}

\section{Proving that $\bigtriangledown_\mu u_\nu=0$ everywhere implies
ultrastatic spacetime}\label{ultrastatic} Assuming
$\nabla_\mu u_\nu=0$, the Frobenius condition\supcite{frob} 
$u_{[\mu}\nabla_{\nu} u_{\rho ]}=0$
is trivially
satisfied. This means that there exists (locally) a slice for which $u^\mu$ is
normal. Introducing coordinates such that $t=\textrm{const}$ in every slice and
letting the spatial coordinates follow the congruence connected to
$u^\mu$, the line element takes the form
\begin{equation} g_{\mu\nu}\left[ \begin{array}{ccccc} g_{tt} \ \ \ & 0 \\ 0 & g_{ij} \end{array}
\right].
\end{equation} 
In this particular system $u_\mu=\sqrt{g_{tt}}\
{\delta^t}_\mu$. Then we readily find 
\begin{eqnarray}
\bigtriangledown_\alpha u_\beta &\equiv& \partial_\alpha u_\beta -
\Gamma^\rho_{\alpha \beta} u_\rho\\ &=& ... \\
&=&\frac{1}{\sqrt{g_{tt}}}   \nonumber \\ && \left( {\delta^t}_\beta
(\partial_\alpha g_{tt}) - \frac{1}{2} \left( \partial_\alpha g_{t\beta}
+ \partial_\beta g_{t\alpha} - \partial_t g_{\alpha \beta}  \right)
\right). \qquad 
\end{eqnarray} 
Letting $i$ and $j$ denote general spatial indices and evaluating this
equation for $\alpha=t,i$ and $\beta=t,j$ (there are four different
combinations) we readily find
\begin{eqnarray} 
\partial_\mu g_{tt}&=&0 \\ 
\partial_t g_{ij}&=&0.
\end{eqnarray} 
Thus in these particular coordinates, choosing a
$t$-labeling such that $g_{tt}=1$, the metric takes the form
\begin{equation} g_{\mu\nu}= \left[ \begin{array}{ccccc} 1 \ \ \ &  0   
        \\ 0  &    g_{ij}({\bf x}) \end{array} \right]. 
\end{equation} 
A spacetime where the metric can be put in this form is called
ultrastatic. Thus $\bigtriangledown_\mu u_\nu=0$ implies an
ultrastatic spacetime. The converse, choosing the preferred observers
in the ultrastatic spacetime as observers, is also obviously true.


\begin{thebibliography}{99}
\bibitem{metric}
\eq{hhh} defines a so
called distance function, or a metric. It can also be used considering events where 
$dx>c dt$. Then $d\tau^2$ is negative which simply means that it is related
to spatial distance rather than temporal distance. 
A distance function like \eq{hhh} corresponds to a flat spacetime, but
see \eq{sc} for an example of a distance function corresponding to a
curved spacetime.
\bibitem{noblack} {Assuming that there is no black hole inside the
crust.}
\bibitem{ingemar} {As was pointed out to me by Ingemar Bengtsson, this
relation is also used by Hawking \& Ellis (\rcite{hawking}), although for
completely different purposes than those of this article.}
\bibitem{raise} {If we had instead considered a metric of the form
$-g^{\mu\nu}+\alpha   u^\mu u^\nu$, where $\alpha$ is some general
number, the inverse would have been $-g^{\mu\nu}+\alpha/(\alpha-1)  
u^\mu u^\nu$. It is only in the case $\alpha=2$ that we can simply raise
the indices of the absolute metric with the original metric to get the
inverse of the absolute metric.}
\bibitem{gamma} The gamma factor is defined as $\gamma=(1-v^2)^{-\frac{1}{2}}$
where $v$ is the relative velocity. With this definition, it follows
from \eq{hhh} that $d\tau=\frac{dt}{\gamma}$.
\bibitem{killdef}{A geometry has a Killing symmetry if there exists a
vector field (called a Killing field) $\xi^\mu$ such that when we shift
our coordinates $x^\mu \rightarrow x^\mu + \epsilon \xi^\mu$ -- the
metric has the same form. As an example we can consider a geometry that
can be embedded as a curved surface. Then there exists a Killing
field directed around the surface (in the azimuthal direction) with a
length proportional to the embedding radius. Also, if there are
coordinates where the metric is independent of one coordinate, then
there is a Killing symmetry with respect to that coordinate. }
\bibitem{energyconditions} {This may well violate the energy
conditions.}
\end{thebibliography}
\end{document}